\documentclass[groupedaddress, aps, preprintnumbers,12pt]{revtex4}

\usepackage{graphicx,amsmath,amssymb}

\newcommand{\bq}{\begin{equation}}
\newcommand{\eq}{\end{equation}}
\newcommand{\bqa}{\begin{eqnarray}}
\newcommand{\eqa}{\end{eqnarray}}
\newcommand{\ben}{\begin{enumerate}}
\newcommand{\een}{\end{enumerate}}
\newcommand{\bc}{\begin{center}}
\newcommand{\ec}{\end{center}}
\newcommand{\bqb}{\begin{eqnarray*}}
\newcommand{\eqb}{\end{eqnarray*}}

\def\gsim{\gtrsim}
\def\lsim{\lesssim}

\def\eg{{\it e.g. }}

\def\etal{{\it et al.}}

\def\swd{s^2_W}

\def\V{ {\cal V }}
\def\A{ {\cal A }}
\def\R{ {\cal R }}

\begin{document}

\preprint{DO-TH 06/03}

\title{Coherent pion production by neutrino scattering off nuclei.}
\author{A. Kartavtsev}
\email[Email: ]{akartavt@het.physik.uni-dortmund.de}
\author{E. A. Paschos}
\email[Email: ]{paschos@physik.uni-dortmund.de}
\affiliation{Universit\"at Dortmund, Institut f\"ur Physik,
D-44221 Dortmund, Germany}
\author{G.J. Gounaris}
\email[Email: ]{gounaris@physics.auth.gr}
\affiliation{Aristotle
University of Thessaloniki, Department of Theoretical Physics,
Gr-54124 Thessaloniki, Greece.}

\begin{abstract}
The main part of coherent pion production by neutrinos on nuclei
is essentially determined by PCAC, provided
that  the leptonic momentum transferred square $Q^2$ remains sufficiently small.
We give the  formulas for the charged and
neutral current cross sections,  including  also the small non-PCAC transverse
current contributions and taking into account the effect of the $\mu^-$-mass.
Our results are  compared with  the
experimental ones and other theoretical treatments.
\end{abstract}
\maketitle

\section{Introduction}
Coherent production of pions by neutrinos has been studied
by many experimental groups and measurements have been made
for neutrino energies ranging from $2$ to $80$ GeV
\cite{Faissner:1983ng,Marage:1984zy,Isiksal:1984vh,Marage:1986cy,
Grabosch:1985mt,Allport:1988cq,Willocq:1992fv,Vilain:1993sf}.
The main characteristics of such  cross sections
is that   the  energy of the recoiling nucleus and  the
 invariant momentum transfer  to it, always remain very small.
A characteristic signature of these events is a sharp peak
in the low $|t|$ region.
In addition to this, all experiments have observed
that the momentum transfer  from  the leptonic sector $Q^2$
also remains very small,  sharply peaking at   $Q^2 \lsim 0.2$ GeV$^2$;
while the dependence of the cross section
on the neutrino energy appears logarithmic at high   energies.

However, problems with the existence of the coherence phenomenon
might appear  at the lower energies used for
 the new oscillation experiments    K2K,   MiniBoone, MINOS etc.
 In particular,  a new measurement by the K2K group
  at an average neutrino energy $E_1=1.3$ GeV,
   has set an upper bound on  the coherent pion production
   by neutrinos, which is  far below the theoretical expectations
\cite{Hasegawa:2005td}. This has raised  questions on how accurately the  coherent
cross section can be calculated  in such a  low energy region,
and whether  detail  event distributions may be predicted.

Theoretical calculations on the other hand,
have  presented general arguments based on the partial conservation of the axial
current (PCAC) and the dominance
of  the axial current by pions or axial vector mesons
\cite{Rein:1982pf,Paschos:2003hs},
or more complicated structures \cite{Belkov:1986hn, Kopeliovich:2004px,
Gershtein:1980vd, Komachenko:1983jv}, occasionally using nuclear physics models
\cite{Kelkar:1996iv}. The situation is not yet settled.

In some  models one starts with the Adler relation \cite{Adler:1964yx}
in the $Q^2= 0$ limit and extrapolates it to small $Q^2$ values. In the work
of Rein and Sehgal \cite{Rein:1982pf} the pole due to  the $a_1(1260)$ resonance is introduced
together with other assumptions for estimating the pion-nucleus cross section.
In several articles, Kopeliovich  \etal  \cite{Kopeliovich:2004px}
have claimed that the pion pole term acting on the leptonic current gives
a small contribution proportional to the lepton mass, and  they are led to argue
that the axial current must be dominated
by heavy meson fluctuation like  $a_1(1260)$  or the
$\rho\pi $ branch point.

Instead, we show here that a careful PCAC
treatment  determines the dominant terms in a unique way.
More specifically, we decompose the leptonic current contribution into a spin=0 and spin=1
state with three helicity components. The inner product of the helicity zero polarization
vector with the axial hadronic current leads to  matrix elements in
the $Q^2\ll \nu^2$ region, determined by PCAC as   $f_\pi T(\pi N\to \pi N)$,
with $T$ being the amplitude for the coherent  pion-nucleus scattering,
which  is a smooth function of $Q^2$, having  no pion pole.
This way, a Goldberger--Treiman--type relation is obtained, determining the
true dominant contribution to coherent neutrino-pion
production. In addition to this, there exist
of course  contributions arising from the
transverse (off shell) vector and axial states,
which are estimated phenomenologically and  turn out to be  very  small.

Since the kinematics for the charged current (CC) cross sections
obey   $Q^2_{\rm min}\sim m_\mu^2\sim m_\pi^2$,
all mass terms are retained in the
calculation of the density matrix of the leptonic current and
the phase space.
For the neutral current (NC) reactions, the neutrino masses are of course
negligible and the formulae are simplified. Using these,
 we plot $d\sigma/dQ^2$ for  NC and CC pion production  at  small $Q^2$,
 and compare the results.

The purpose of the present paper is to contribute in clarifying the theoretical
framework for coherent pion production processes.
More explicitly, we show that for energies of the produced
pion above a few GeV,
the main contribution to the coherent neutrino-pion
 production  is   determined by  PCAC and the
 pion-Nucleus coherent scattering data. The remaining contributions
 arising   from transverse off-shell
 vector and axial mesons, must always be very small. In particular, the
  transverse vector contribution is expressed in terms of the $\pi^0$ coherent
  photoproduction data, and it is thus reliably estimated.
  Estimating the axial transverse  contribution is  more difficult,
  but a  Regge analysis indicates that
  it should be comparable or probably smaller than the transverse vector
  contribution.

In the following,  we present in Section II
the general formalism   for coherent $\pi^\pm$  or $\pi^0$
 production through neutrino scattering off a nucleus.
 In Section III we describe the experimental data  and present
 our numerical results. The Conclusions appear in Section IV.

\section{The Formalism}
 For clarity, we first concentrate
on  the  $\pi^+$ production process  through  coherent $\nu_\mu$
scattering off a heavy nucleus $N$, according to the process
\bq
\nu_\mu (k_1)~ N(P) \to \mu^-(k_2)~ \pi^+(p_\pi)~ N(P')~~,~~\label{CC-process}
\eq
where the momenta are indicated in parentheses. Here  $q=k_1-k_2$ is the momentum
four-vector
transferred from the leptonic current to the nucleus N, so that its energy-component
$\nu=q_0=E_1-E_2$ (with $E_1$ and $E_2$ being   the
$\nu_\mu$ and $\mu^-$ laboratory energies respectively)
denotes the energy given by the  current to the
$\pi^+ N$-pair in the Lab frame.

In the coherent scattering regime  the nucleus spin is not flipped, and
its recoil must be minimal, so that
$\nu\simeq E_\pi $, with
 $E_\pi$ being  the pion energy in the Laboratory frame.
 The existing experimental data also  suggest that in the coherence  regime
 $0\leq Q^2=-q^2 \lsim 0.2 ~{\rm GeV^2}$, and that
 the squared   momentum-transfer in the hadronic system
   $t=(q-p_\pi)^2=(P-P')^2$ is  peaked at very small values.

The invariant amplitude for the process (\ref{CC-process}) may  be written
as
\bq
T_W = -\frac{G_FV_{ud}} {\sqrt{2}}\bar u(k_2)\gamma^\rho (1-\gamma_5)u(k_1)
(\V^+_\rho-\A^+_\rho) ~~,~~ \label{CC-amplitude}
\eq
where the first factor gives  the $(\nu_\mu\to \mu)$-matrix element of the
leptonic current, while
\bq
 \V^+_\rho= \langle \pi^+N| V^1_\rho+i V^2_\rho|N\rangle ~~,~~
  \A^+_\rho= \langle \pi^+N|A^1_\rho+iA^2_\rho|N\rangle ~~,~~ \label{hadronic-elements}
\eq
describe (in momentum space)  the hadronic matrix elements of the
charged vector and axial  currents respectively. $V_{ud}$ in (\ref{CC-amplitude})
denotes the appropriate CKM matrix element.

Since, the charged leptonic current is not conserved  $(m_{\mu}\neq 0)$, it contains
spin=0  degrees of freedom described by its component along the vector
\bq
\epsilon^\rho_l=\frac{q^\rho}{\sqrt{Q^2}}~~, \label{eps-scalar}
\eq
 as well as spin=1 degrees of freedom describing
   off-shell gauge bosons with the   helicity polarization vectors
\bq
\epsilon^\rho (\lambda=\pm 1) =
\mp \left (
\begin{matrix}
0 \cr  1 \cr  \pm  i \cr 0
\end{matrix}  \right )^\rho ~~,~~
\epsilon^\rho (\lambda=0) =
\frac{1}{\sqrt{Q^2}} \left (
\begin{matrix}
|\vec{q}|  \cr 0   \cr 0    \cr q_0
\end{matrix}  \right )^\rho , \label{eps-vector}
\eq
 when  $\vec{q}$ is  taken along the  $\hat z$-axis.
The  $\lambda=\mp 1$ polarizations  in (\ref{eps-vector}) are often denoted as L(R)
 respectively,   the vanishing helicity  vector $\epsilon^\rho (\lambda=0)$
 is identical to  $\epsilon_S^\rho$ of    \cite{Bjorken:1969in},
 and  $\epsilon^\rho (\lambda)q_\rho=0$
 is of course always satisfied.

 Anticipating that we later  integrate over all  relative angles between
 the $(\vec{k}_1,\vec{k}_2)$-leptonic plane and the $(\vec{q},\vec{p}_\pi)$ pion production
  plane,  the only density matrix elements needed for the above spin=0 and 1 states
  hitting the nucleus N are \footnote{Since $Q^2>0$,
  the $\lambda=0$ polarization vector in
 (\ref{eps-vector}) is time-like,
 while all other vectors in (\ref{eps-vector}, \ref{eps-scalar}) are space-like.
 Because of this the form of the closure condition becomes
$$ \sum_{\lambda=0,\pm 1} (-1)^\lambda \epsilon^\mu(\lambda)
 \epsilon^{\nu *}(\lambda) -\epsilon^\mu_l\epsilon^\nu_l=g^{\mu\nu}$$
 which is  important   in determining    the sign of the $\tilde L_{l0}$
 term in (\ref{density-matrix}).}
\bqa
\frac{(\tilde L_{RR}+\tilde L_{LL})}{2}&=&Q^2
\Big [1+\frac{(2E_1-\nu)^2}{\vec{q}^2}\Big ]
-\frac{m_\mu^2}{\vec{q}^2}[2 \nu (2 E_1-\nu)+m_\mu^2]~~,
\nonumber \\
\frac{(\tilde L_{RR}-\tilde L_{LL})}{2}&=&
-\frac{2 [Q^2 (2E_1-\nu)-\nu m_\mu^2]}{|\vec{q}|}~~,
\nonumber \\
\tilde L_{00}&=& \frac{2 [Q^2 (2E_1-\nu)-\nu m_\mu^2]^2}{Q^2 \vec{q}^2}
-2(Q^2+m_\mu^2)~~,
\nonumber \\
\tilde L_{ll}&=& 2 m_\mu^2 \Big(\frac{m_\mu^2}{Q^2}+1 \Big )~~,
\nonumber \\
\tilde L_{l0}&=& \frac{2 m_\mu^2 [Q^2 (2E_1-\nu)-\nu m_\mu^2]}{Q^2 |\vec{q}|}~~.
\label{density-matrix}
\eqa

Using these  and
 the hadronic current  elements in (\ref{hadronic-elements}),
 the square of the amplitude in (\ref{CC-amplitude}), summed over all
 $\mu^-$ polarizations,  is written as
\bqa
\overline{|T_W|^2}&=&  G_F^2|V_{ud}|^2\Bigg \{
\frac{(\tilde L_{RR}+\tilde L_{LL})}{2}
\sum_{\lambda=L,R}|(\V^+-\A^+)\cdot \epsilon (\lambda)|^2
\nonumber \\
&+ &\frac{(\tilde L_{RR}-\tilde L_{LL})}{2}
\Big [|(\V^+-\A^+)\cdot \epsilon (R)|^2-|(\V^+-\A^+)\cdot \epsilon (L)|^2 \Big ]
\nonumber \\
&+ & \tilde L_{00} |(\V^+_\rho-\A^+_\rho) \epsilon^\rho(\lambda=0) |^2
+\frac{\tilde L_{ll}}{Q^2} |(\V^+_\rho-\A^+_\rho) q^\rho |^2
\nonumber \\
&+ & \frac{2\tilde L_{l0}}{\sqrt{Q^2}}
\Re \Big ([(\V^+_\rho-\A^+_\rho) \epsilon^\rho(0)]
\cdot [ (\V^+_\mu-\A^+_\mu) q^\mu]^*\Big )\Bigg \}~~, \label{Tw-squared}
\eqa
where the first two terms may  be interpreted as
  giving  the contributions from the transverse spin=1
components of the hadronic currents,   the third term gives
 the helicity $\lambda=0$  hadronic contribution,
 the fourth term  arises from the spin=0 component,
and finally the last term from the interference of the latter two.

We first concentrate on the axial current matrix elements
  in the last three terms of (\ref{Tw-squared}), which
    turn out to give the most important contributions,
    for the GeV-scale kinematic region where coherence is relevant.
  The  pion poles    contained   in these terms, induce  a singularity at low
  $Q^2$, which   must be   carefully  separated, before any approximation   is made.

To achieve this we note that  the axial hadronic
  element in (\ref{hadronic-elements}) consists of the pion pole contribution, and
  the rest we call $\R^\rho$, induced by $a_1(1260)$ and any other isovector
  axial meson that might exist. It  is thus, written as
  \bq
  -i  \A^+_\rho=  \frac{f_\pi \sqrt{2} q^\rho}{Q^2+m_\pi^2}T(\pi^+N \to \pi^+N)-
  \R^\rho~~,  \label{PCAC-rel1}
  \eq
where $T(\pi^+N \to \pi^+N)$ is the $\pi$-nucleus purely hadronic
invariant amplitude, $f_\pi\simeq 92 MeV$,  and
$\R^\rho$ is a very smooth  function of $Q^2$
whose dependence on it is  ignored \footnote{In principle we could insert here  some
pole contribution from the $a_1(1260)$ axial vector boson, in order to describe
possible $Q^2$ dependence in $\R^\rho$; but this resonance is so far away from the
relevant $Q^2$ region, that such an effort does not seem useful.}.
The usual PCAC treatment  leads to
\bqa
  && -i q^\rho \A^+_\rho=  \langle \pi^+N|\partial^\rho A^+_\rho|N\rangle =
   \frac{f_\pi m_\pi^2 \sqrt{2}}{Q^2+m_\pi^2}~T(\pi^+N \to \pi^+N)
   \nonumber \\
 & & ~~~~~~~~~ =  - \frac{f_\pi Q^2 \sqrt{2}}{Q^2+m_\pi^2}~T(\pi^+N \to \pi^+N)
   -q_\mu\R^\mu    ~~,  \label{PCAC-rel2} \\
   && \hspace{0cm}  \Rightarrow ~~~~~
   q^\mu \R_\mu= -f_\pi \sqrt{2}~ T(\pi^+N \to \pi^+N) ~~. \label{PCAC-rel3}
 \eqa
It is amusing to emphasize that  (\ref{PCAC-rel3}) is strongly reminisced of the
classical Goldberger--Treiman treatment, where the pion pole not only
determines $\partial^\mu A_\mu$, but in fact also the complete axial current
coupling \cite{Goldberger}.

Using now (\ref{PCAC-rel1}), and
 $\epsilon(0)_\rho q^\rho=0$ implied by
(\ref{eps-scalar}, \ref{eps-vector}), we conclude
\bq
\epsilon(0)^\rho \A^+_\rho=- i\epsilon(0)^\rho \R_\rho \simeq
i \frac{f_\pi \sqrt{2}}{\sqrt{Q^2}}~T(\pi^+N \to \pi^+N)~~, \label{PCAC-rel4}
\eq
where  in the first step the pion pole contribution vanishes identically,
while the last step is due to the smoothness of $\R^\rho$
and the restriction to  $\nu \gg \sqrt{Q^2}$, which  justify
the  approximation
\bq
\nu \gg \sqrt{Q^2}
\quad \Rightarrow \quad
\epsilon^\rho(0)\simeq q^\rho/\sqrt{Q^2}~~. \label{ap-helicity0}
\eq
In order for the simple pion dominating picture
obtained below to be  valid,
the kinematics should always be chosen such that
 $\nu \gg \sqrt{Q^2}$.  To guarantee this we
 introduce the parameter $\xi$ in (\ref{nu-limits}) of the Appendix.

Since (\ref{PCAC-rel4}) is our most important theoretical result,
it might be  worth  emphasizing that it would be incorrect
to apply  the  approximation
(\ref{ap-helicity0}) directly on the $\epsilon(0)^\rho \A^+_\rho$
computation using  (\ref{PCAC-rel1}),
because that    will  replace the identically vanishing expression
$\epsilon^\mu(0)q_\mu/(Q^2+m_\pi^2)$,
 by  the non-vanishing and in fact
large quantity $-\sqrt{Q^2}/(Q^2+m_\pi^2)$
\footnote{ See \eg also at C. Itzykson and J.-B. Zuber,
  Quantum Field Theory, McGraw-Hill 1980,
  p.535, particularly the remark immediately after equation (11-113).}.

The relations (\ref{PCAC-rel2}, \ref{PCAC-rel4}) fully determine the axial current
contribution to the last three terms of (\ref{Tw-squared}).
We also remark that  these results are consistent
with the Adler theorem in the parallel lepton configuration
 \cite{Adler:1964yx, Adler:2005ad}, provided we set $m_\mu=0$.

Furthermore, the vector hadronic elements
in the last three terms of (\ref{Tw-squared})   give no contribution;
since the vector current is conserved,   and the applicability of
(\ref{ap-helicity0}) for  calculating $\epsilon^\rho(\lambda=0)\V^+_\rho$
is  guaranteed  by the absence of
any low mass singularity.
Moreover, since in the coherence regime there is  no  $R-L$
polarization sensitivity to the vector or axial-vector boson cross sections,
there will not be  any contribution from the second term
in (\ref{Tw-squared}).

Thus,  the CC neutrino coherent pion production cross section off a nucleus N
 becomes
\bqa
&& \frac{d\sigma(\nu N\to \mu^-\pi^+N)}{dQ^2 d\nu dt}=
\frac{G_F^2 |V_{ud}|^2 \nu }{2(2\pi)^2  E_1^2  } \Bigg \{
\nonumber \\
&&  \frac{f_\pi^2}{Q^2}\Big [\tilde L_{00}+ \tilde L_{ll}
\Big (\frac{m_\pi^2}{Q^2+m_\pi^2} \Big )^2
+2 \tilde L_{l0} \frac{m_\pi^2}{Q^2+m_\pi^2}  \Big ]
 \frac{d\sigma(\pi^+ N\to \pi^+ N)}{dt}
\nonumber \\
&& + \frac{(\tilde L_{RR}+\tilde L_{LL})}{2}
\Big [\frac{1}{2\pi \alpha} \frac{d\sigma( \gamma  N\to \pi^0 N)}{dt}
  +\frac{d\sigma(A^+_T N\to \pi^+N)}{dt} \Big ]
\Bigg \}~~,
\label{dsigma-full}
\eqa
expressed in terms of the leptonic density matrix elements
 in (\ref{density-matrix}). In deriving this expression
we have integrated over all  angles between the lepton-
and  $(\vec{q},\vec{p}_\pi)$-planes, and ignored
any Vector-Axial interference in (\ref{Tw-squared}),
since it will anyway cancel out after the
 $t$-integration we  do,  before comparing to the
 experimental data. Notice that in contrast to (\ref{Tw-squared}),
 the presentation  in (\ref{dsigma-full})
 first gives the numerically most important terms arising from the
  $\lambda=0$ and the spin=0 components of the leptonic current, and
  then the less important contributions from its  transverse vector
  and axial components.

 In treating the phase space in
 (\ref{dsigma-full}) we have used
\bq
 W^2 \equiv (q+P)^2=M_N^2-Q^2+2 M_N\nu \simeq M_N^2+2M_N \nu\simeq
 M_N^2+2M_N E_\pi ~~, \label{W-coherence}
\eq
for the invariant mass-squared of the $\pi^+N$-pair.  Thus,
 the three differential cross sections occurring in the r.h.s. of
(\ref{dsigma-full}) should be thought as functions
of  $t$, and the laboratory energy $\nu \simeq   E_\pi$, of the
particle hitting  the nucleus.

We next turn to the last  two terms within the curly brackets in (\ref{dsigma-full}),
which are induced by  the transverse components of all  off-shell vector and axial
vector mesons coupled to the $\V^+_\nu$ and $\A^+_\nu$ matrix elements at very small $Q^2$;
compare (\ref{hadronic-elements}).
The  vector term  is directly related,
(after an isospin rotation producing a factor 2), to  $\pi^0$
photoproduction  for  unpolarized photons.
In deriving this, it is important to realize that
the isoscalar part of the electromagnetic
current does not contribute to the coherent $\pi^0$ amplitude.
This contribution is  estimated in the next section,  using the experimental data
\cite{Krusche}.

The transverse  axial  term within the curly brackets
 in (\ref{dsigma-full})
\bq
\frac{d\sigma(A^+_T N\to \pi^+N)}{dt}=
\frac{\sum_{\lambda=L,R}|\A^+\cdot \epsilon (\lambda)|^2}{128 \pi \nu^2 M_N^2}~~,
\label{axial-dsigma}
\eq
 expressed in terms of the axial matrix element
 of (\ref{hadronic-elements}),  describes the cross section  for $\pi^+$-production
through     "transversely polarized charged axial currents".
To calculate it, we would need  to know   all possible
$a_1^+(1260)$-type mesons that couple to the axial  current, their  couplings to it, and
the corresponding $\sigma(a_{1T}^+ N \to \pi^+ N)$ off-shell transverse $a_1$ cross sections,
at very small $Q^2$. We  estimate this also in the next section.

A similar procedure may be carried out
for the NC coherent $\pi^0$-production, for which the result
\bqa
&& \frac{d\sigma(\nu N\to \nu \pi^0 N)}{dQ^2 d\nu dt}=
\frac{G_F^2 \nu }{4(2\pi)^2 E_1^2  } \Bigg \{
 \frac{f_\pi^2}{Q^2} \tilde L_{00}
 \frac{d\sigma(\pi^+ N\to \pi^+ N)}{dt}
\nonumber \\
&& + \frac{(\tilde L_{RR}+\tilde L_{LL})}{2}
\Big [\frac{(1-2\swd)^2}{2\pi \alpha} \frac{d\sigma( \gamma  N\to \pi^0 N)}{dt}
  +\frac{d\sigma(A^+_T N\to \pi^+N)}{dt} \Big ]  \Bigg \}~~
\label{dsigma-NC-full}
\eqa
is found, provided the assumption
\bq
\frac{d\sigma(\pi^+ N\to \pi^+ N)}{dt}\simeq \frac{d\sigma(\pi^0 N\to \pi^0 N)}{dt}
~~, \label{piplus0-sym}
\eq
is made, which  is on the same footing as the isospin rotation
we used in writing (\ref{dsigma-full}) in terms of the
$\pi^0$ photoproduction data.

In (\ref{dsigma-NC-full}), the
 leptonic density matrix elements are given by
the same expressions as in (\ref{density-matrix}), with the obvious substitution
$m_\mu \rightarrow 0$. Comparing
the NC  result (\ref{dsigma-NC-full}), to the CC in (\ref{dsigma-full}),
 we see that there is no CKM factor  now, and that
the axial contribution to the NC cross section is a  factor 2 smaller than
the CC one. For the vector contribution though, an extra reduction
by a factor $(1-2\swd)^2$ appears, which is due to the fact that Z couples  not only
to the $SU(2)_L$-current, but also to the isovector part of
the electromagnetic current.

\section{Numerical Estimates and  Results}
For numerical estimates we must calculate the three cross sections
appearing in equations
(\ref{dsigma-full}) and  (\ref{dsigma-NC-full}). The dominant contribution comes
from $\sigma (\pi^+ N\rightarrow \pi^+N)$, for which we  use data on coherent
scattering of pions on nuclei. This being the dominant term, we  calculate
it precisely and present the results in the figures below.
The other two cross sections involve coherent photoproduction
of pions and the  $A^+_T N\rightarrow \pi^+N$ process,  where the axial vector
particles are transversely polarized and give smaller
contributions. We have estimated them  using available data
and showed  that they are  very small. Thus,
assigning to the latter two
cross section an uncertainty even as large as 50\%,  does  not
affect our results.

For isoscalar targets,  like $C^{12},~O^{16}$..., isospin symmetry implies
$d\sigma(\pi^+ N)\simeq d\sigma(\pi^- N)\simeq d\sigma(\pi^0 N)$.
In the actual calculation we use the coherent pion-Carbon scattering data
\cite{Binon:1970ye}, with additional data being available on other nuclei and
other energies in  \cite{Kahrimanis:1997if, Scott:1972yh}.
Data from other nuclei are   normalized to Carbon,
using  the A-dependence law $A^{2/3}$,
observed in hadronic experiments \cite{Ashery:1981tq} and
 in pion photoproduction  \cite{Meyer}.

\begin{figure}[h!]
\includegraphics[width=0.73\textwidth]{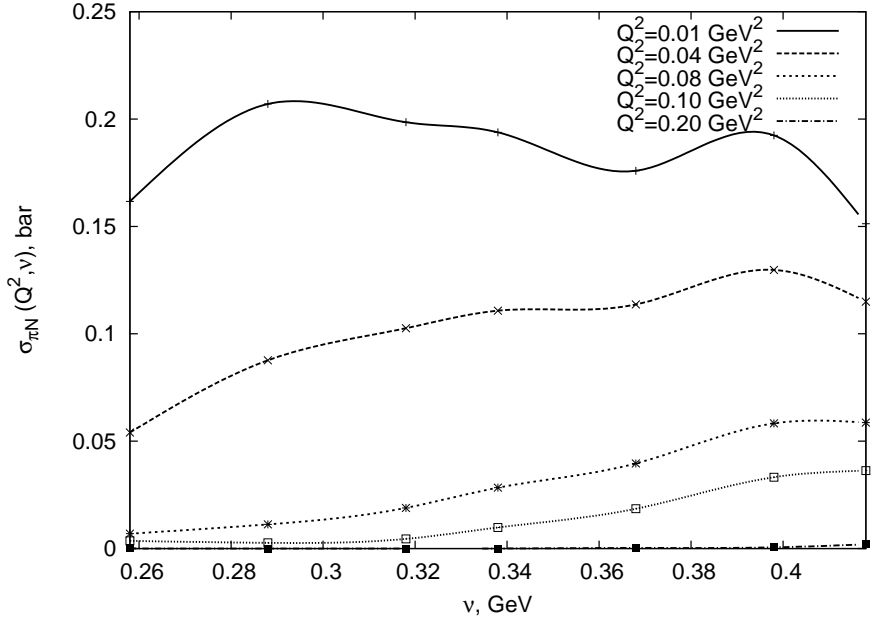}
\caption{\label{pioncrs} Pion-Carbon cross section integrated over $t$ in the
range discussed in the text, as a function of $\nu$,
at different values of $Q^2$.}
\end{figure}

In all cases,  $\nu$ is identified with the laboratory pion energy, and the integration
  over $t$ is done for the low values shown in Appendix A.
   Thus, the  pion-carbon cross section
$d\sigma(\pi^+C\rightarrow\pi^+C)/dt$ is integrated  from
$|t|_{min}$ given in  (\ref{tmin}),
to $|t|_{max}\simeq 0.05$  GeV$^2$  corresponding  to the first dip of
the pion-carbon cross section. We checked that this  $t_{min}$ is
sufficiently large for the cross section  to be outside
the Coulomb peak  which also contributes to the $\pi^\pm N$  scattering
at very small angles \cite{Binon:1971ye}.
The resulting integrals are functions of
the pion energy  $\nu$, and
the momentum transfer squared $Q^2$, introduced through the lower limit of
the $t$-integration.
The results  are shown in Fig.\ref{pioncrs}, for various  $\nu$ and $Q^2$ values.

\begin{figure}[h!]
\includegraphics[width=0.73\textwidth]{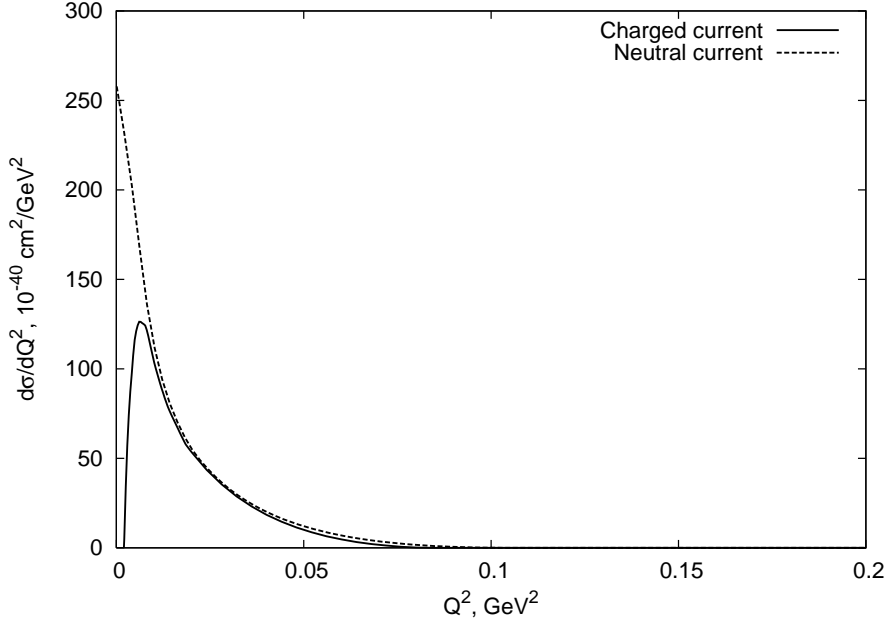}
\caption{\label{diff} The differential cross sections
for  the coherent pion production by neutrinos
$d\sigma(\nu N\to \mu^-\pi^+N)/dQ^2$  and
$d\sigma(\nu N\to \nu \pi^0N)/dQ^2$, for $E_1=1.0~ {\rm GeV}$. Only
$\sigma (\pi^+ N\rightarrow \pi^+N)$ has been taken into account,
 since the transverse vector and axial contribution are negligible.
The curves correspond to $\xi=3$; see (\ref{nu-limits}).}
\end{figure}

Integrating next  (\ref{dsigma-full}, \ref{dsigma-NC-full} )
over $\nu$ in the range of (\ref{nu-limits}),
we  obtain the $\sigma (\pi^+ N\rightarrow \pi^+N)$  contribution to the
differential cross sections
 $d\sigma(\nu N\to \mu^-\pi^+N)/dQ^2$  and
 $d\sigma(\nu N\to \nu \pi^0N)/dQ^2$, for the  CC and NC reactions
depicted in  Fig.\ref{diff}. We notice that the shapes of the CC and
the NC distributions are different, most notably
 because of the $m_\mu$-mass effects.
 The results in Fig.\ref{diff} correspond to $\xi=3$,
 defined in (\ref{nu-limits}), in order to be  consistent
 with (\ref{ap-helicity0}).  We also note that such shape differences as indicated in
 Fig.\ref{diff},  must be taken into account,
 in the comparison with the Adler parallel configuration.

\begin{figure}[h!]
\includegraphics[width=0.5\textwidth,angle=-90]{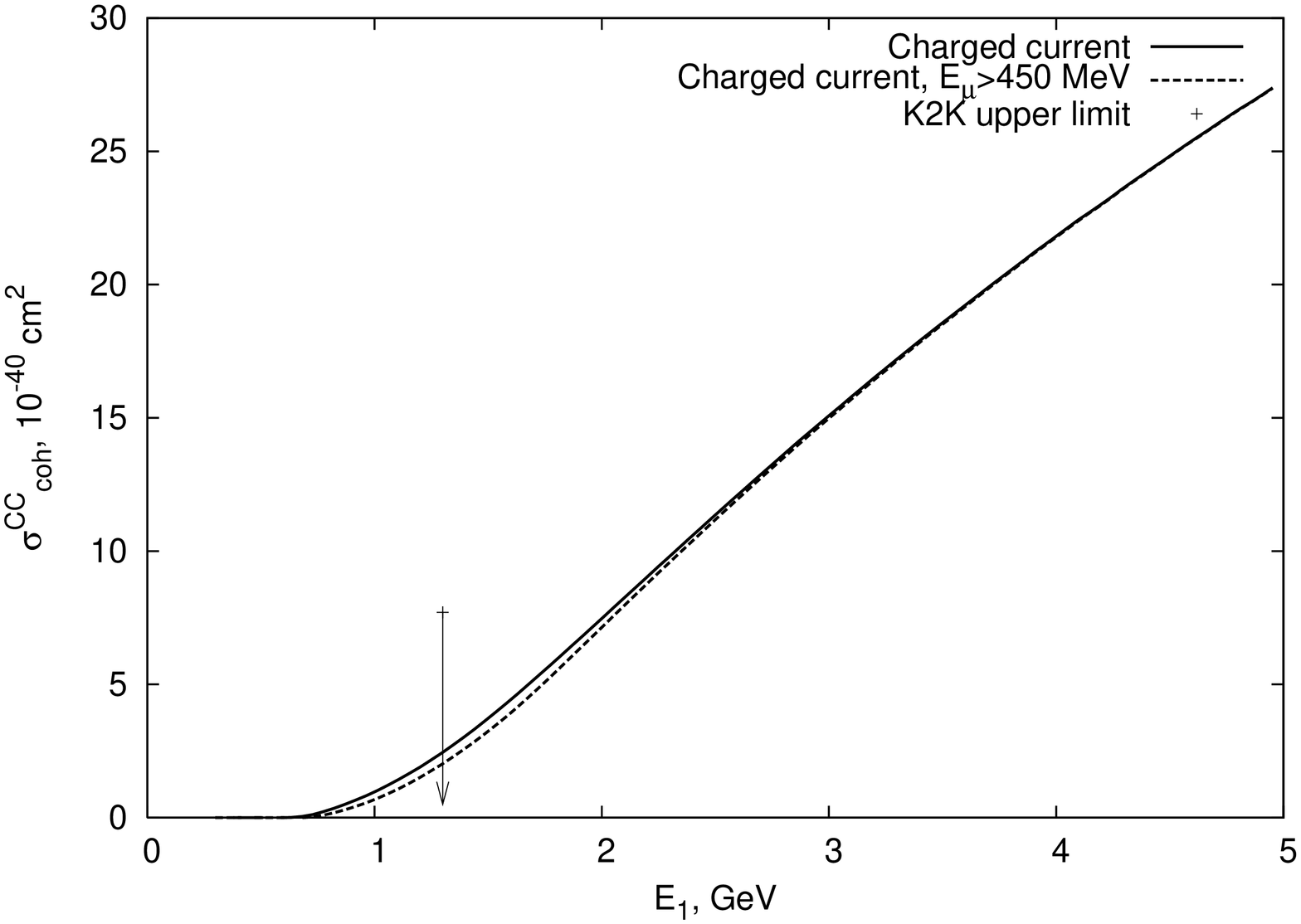}
\caption{\label{totalCC}Integrated cross section  of the coherent pion production
per carbon nucleus  by neutrinos in the CC case.
Only   $\sigma (\pi^+ N\rightarrow \pi^+N)$ has been taken into account,
  since the transverse vector and axial contribution are  negligible.
  The upper bound is from K2K including one standard
deviation. Dotted line  represents integrated cross section with a
threshold value for the muon energy $E_\mu>450$ MeV.
The theoretical curves correspond to $\xi=3$; compare(\ref{nu-limits}).}
\end{figure}

\begin{figure}[h!]
\includegraphics[width=0.5\textwidth,angle=-90]{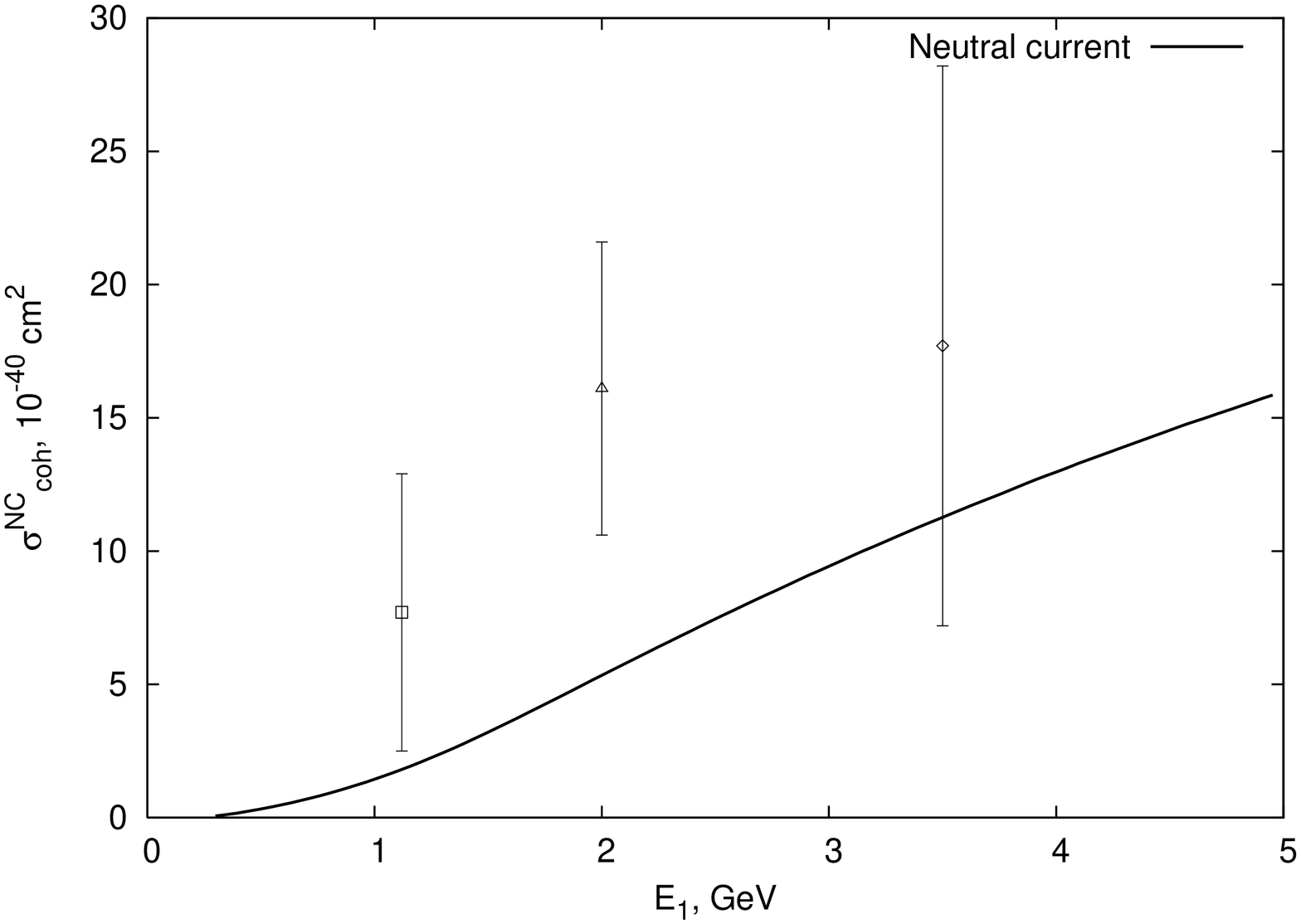}
\caption{\label{totalNC}Integrated cross section  of the coherent pion production
per carbon nucleus  by neutrinos in the NC case.
Only   $\sigma (\pi^+ N\rightarrow \pi^+N)$ has been taken into account,
  since the transverse vector and axial contribution are negligible.
The experimental points for NC are from: $\square$ MiniBoone \cite{Raaf:2005up},
$\triangle$ Aachen-Padova \cite{Faissner:1983ng}, $\diamondsuit$ Gargamelle
\cite{Isiksal:1984vh}. The theoretical curve corresponds to $\xi=3$.}
\end{figure}

Finally, integrating
over $Q^2$ in the region (\ref{Q2-limits}), we obtain the results in
Fig.\ref{totalCC} and Fig.\ref{totalNC} are obtained.

We   next turn to the transverse vector and axial contributions
supplying the terms proportional to the density matrix elements
$ \tilde L_{RR}+\tilde L_{LL}$ in (\ref{dsigma-full}, \ref{dsigma-NC-full}).
For the photon induced reaction, there exist data on the photoproduction of
mesons off nuclei \cite{Krusche, Bartholomy, Meyer}.
The A-dependence reported in \cite{Meyer} is $A^{2/3}$ which indicates
that the same shadowing as in $\pi$-nucleus interactions takes place. Using
then data on Pb from Fig. 9 of \cite{Krusche} at $E_\gamma =200-350$ MeV,
 and integrating them over the first peak, we obtain
\bq
\frac{1}{2\pi\alpha} \int_{|t_{\rm min}|}^{0.01\rm GeV^2} dt
~\frac{d\sigma( \gamma  N\to \pi^0 N)}{dt}
\left (\frac{12}{207}\right )^{2/3} \simeq 1.40 ~mb ~~, \label{photo-data}
\eq
where the  factor $1/2\pi\alpha $ comes from the elimination of the
electromagnetic coupling, and  $(12/207)^{2/3}$ from changing  the cross
section from Lead to  Carbon. The numerical value in (\ref{photo-data})
should be compared with the upper most curve in our Fig.\ref{pioncrs}.
We note that the transverse vector current contribution is approximately  1\%,
of the pion contribution. In addition to it, the ratio of their coefficients
in (\ref{dsigma-full}),  $ (\tilde L_{RR}+\tilde L_{LL})/2$ to
$f_\pi^2 [\tilde L_{00}+...]/Q^2$ in the interesting kinematic region is
$\sim 0.2$. We conclude therefore, that the transverse vector-current contribution
to (\ref{dsigma-full}, \ref{dsigma-NC-full}) is negligible, compared the
pion contribution.

Estimates of the transverse axial current contribution
at low energies are more difficult, because of the absence of  data.
However, as argued below, this  contribution
to (\ref{dsigma-full}, \ref{dsigma-NC-full})
should  be very small and probably  smaller than the
transverse vector  one.

A very rough estimate  for  (\ref{axial-dsigma}) may be obtained  by assuming that
it receives important contributions from  the $a_1^+(1260)$ resonance.
We need two kinds of measurements for this.
The first is the   partial $\tau^-$-width
$\Gamma(\tau^-\to a_1^- \nu_\tau)$,  which determines
 the $a_1$ coupling to the axial current $f_{a_1}$,
defined through (compare (\ref{hadronic-elements}))
\bq
\langle 0|A^1_\rho+i A^2_\rho |a_1^+\rangle = \frac{m_{a_1}^2}{f_{a_1}}
\epsilon_\rho(a_1) ~~, \label{a1-coupling1}
\eq
using
\bq
\Gamma(\tau^-\to a_1^- \nu_\tau)=\frac{G_F^2 m_{a_1}^2m_\tau^3}{16\pi f_{a_1}^2 }
\left (1-\frac{m_{a_1}^2}{m_\tau^2}\right )^2
\left (1+\frac{2 m_{a_1}^2}{m_\tau^2}\right )~~,
\label{a1-coupling2}
\eq
where $(m_{a_1},~\epsilon_\rho(a_1))$ are the $a_1$ mass and polarization vector,
 and $m_\tau$ is the   $\tau$ mass.
Unfortunately the data for $\tau^-\to a_1^- \nu_\tau $ do not show
a clear $3\pi$ resonant state.

Using  as an alternative the corresponding coupling of the $\rho$-meson
to the isovector current $f_\rho^2 \simeq 32$, determined from
\eg the $\Gamma(\rho^0\to e^-e^+)$ data, and taking into account the fact that
the $a_1$-coupling to the axial current could not be stronger \cite{Sakurai},
we expect
\bq
f_{a_1}^2 \gsim 32 ~~. \label{a1-coupling3}
\eq

If in addition some data on $d\sigma(\pi^\pm N \to a_{1T}^\pm N)/dt$ for
\underline{transverse} $a_1$ production were available,
we would estimate
\bq
\frac{d\sigma(A^+_T N\to \pi^+N)}{dt} \sim \frac{2}{f_{a_1}^2}
\frac{d\sigma(\pi^+ N \to a_{1T}^+ N)}{dt}~~,
\label{a1-dsigma}
\eq
where the laboratory energy of the incident  pion is  again  identified with
$\nu$.

To get a feeling of the relative magnitude of the transverse axial  versus transverse
vector contribution, we compare the integrated $\pi^- p\to a_1^- p$ data
at $E_\pi= 16 ~{\rm GeV} $ of \cite{Ballam}, to the $\gamma p\to \pi^0 p$ data
at $E_\gamma =6$ GeV of  \cite{PhysRevD.1.27}.

The integrated diffractive cross section found in \cite{Ballam}
at $E_\pi= 16 ~{\rm GeV} $ is
$\sigma(\pi^- p\to a_1^- p)=250 \pm 50 ~{\rm \mu b}$.
Most of this is of course helicity conserving and refers to the production
of $a_1$ with vanishing  helicity. According to the authors estimate  \cite{Ballam},
 the transverse helicity
part constitutes a fraction of $0.16 \pm 0.08$ of this.
Substituting this  in
(\ref{a1-dsigma}), using  (\ref{a1-coupling3}), we find
\bq
\sigma(A^-_T p\to \pi^- p)~ \lsim ~ 2.5\pm 1.2 ~{\rm \mu b}~~,
\label{a1-sigma-Ballam}
\eq
which should be compared with the transverse vector contribution
\cite{pi0-photoproduction, PhysRevD.1.27}
\bq
\frac{1}{2\pi \alpha} \sigma(\gamma p\to \pi^0 p)\simeq 5 ~{\rm \mu b}
\quad {\rm at} ~~ E_\gamma= 6 ~~{\rm GeV}.
\label{gamma-sigma-Bartholomy}
\eq
In comparing (\ref{a1-sigma-Ballam}, \ref{gamma-sigma-Bartholomy}) we should
remember that the transverse vector and axial processes in (\ref{dsigma-full}),
are both determined by helicity-flip amplitudes.
But in contrast to the $\omega$-Regge trajectory which contributes
uninhibited  to the coherent vector amplitude \cite{Rein:1982pf}; the
only established Regge singularity that can contribute
to the coherent axial amplitude would had been the Pomeron, provided
the associate $a_1$-particles had helicity zero.
Since the currents we consider are transverse
though,  the only possible contributions to
the axial amplitude arises, either from  the small
s-channel helicity violating component of the Pomeron  \cite{SCHC1,SCHC2}, or the
generally  unimportant $\sigma$-trajectory.
On the basis of these, we  conclude that  (\ref{a1-sigma-Ballam}) is very
likely an overestimate. For
coherent production on a Carbon target, we must scale up the proton estimates
(\ref{a1-sigma-Ballam}, \ref{gamma-sigma-Bartholomy})  by
 a factor $12^{2/3}\simeq 5.2$, so that they always remain very small in
 comparison to the pion-Carbon coherent result  that we plotted as the upper most
curve in the  Fig. \ref{pioncrs}.

To sum up, the limited amount of data forced us to use
phenomenological  estimates which imply
that the transverse contributions are very small in comparison
to the pion term. Our
results  in Figs.\ref{pioncrs}-\ref{totalNC},
based on the pion-nucleus data only, can be considered as lower bounds,
with the actual cross sections being  a few percent above them.\\

We  turn next to the implications for
the  oscillation  experiments. Figure \ref{totalCC} shows our results for
CC coherent contribution  to
neutrino-pion production $\sigma^{CC}_{\rm coh}(E_1)$ for $\nu \geq  \xi\sqrt{Q^2}$
for $\xi=3$. The value of $\xi=3$ is chosen so that condition (\ref{ap-helicity0})
is satisfied.
The figure  shows an almost linear increase with neutrino energies.
We note that there is a
rapid growth of the cross section, up to $E_1\sim 5$ GeV. In fact
at $E_1=2.0$ GeV the cross section is almost three times bigger
than   at 1.0 GeV.
For $E_1=1.3$ GeV and $\xi=3$, the predicted coherent  CC
cross section on carbon target is $\sigma^{CC}_{\rm coh}=2\times  10^{-40}$cm$^2$
with the $E_2 \equiv E_\mu>450$ MeV cut applied.

Unfortunately there exist no experimental data that take into account
the $\xi=3$-cut we have imposed for consistency with
our approximation (\ref{ap-helicity0}). The only existing data
for coherent pion production on Carbon \cite{Hasegawa:2005td},
\bq
\sigma^{CC}_{\rm coh} \lsim (7.7\pm 1.6\,{\rm  (stat) }
\pm 3.6 \,{\rm (syst)})\cdot 10^{-40}~ {\rm cm^2}~~ \label{Hasegawa-res}
\eq
are obtained by integrating over all $\nu$-values larger than $\nu_{min}$
appearing in (\ref{numin}), and therefore provide an upper bound to the value
  obtained when  the $\xi=3$-cut is   imposed. They are
of course consistent with our result \footnote{We mention
that for $\xi=1$ the cross section is substantially
bigger. For $E_1=1.3$ GeV it is 7.6$\times 10^{-40}$ cm$^2$,
which is still smaller than the experimental upper bound
(\ref{Hasegawa-res}).  At  higher values of $E_1$ the cross section is
approximately 1.5 to 2 times bigger.}.

Finally, we apply our work to the coherent production of $\pi^0$ in neutral
current reactions. This reaction is an important background in oscillation experiments
searching for the oscillation of $\nu_\mu$'s to $\nu_e$'s.
Several oscillation experiments use two detectors with a long--base--line.
The far away detector searches
among other channels also
for $\nu_e\rightarrow e^-$ interactions. The
$\pi^0$s produced via coherent scattering decay to two photons whose Cherenkov
light mimics that of electrons. Furthermore, when the oscillation is to other types
of active neutrinos all species contribute equally to coherent scattering, but only
$\nu_e$'s produce electrons through the charged current. Thus a good
understanding of coherent $\pi^0$ production is very important.

The NC cross section is calculated from
(\ref{dsigma-NC-full}),  assuming $\sigma(\pi^0C\rightarrow \pi^0
C)\simeq\sigma(\pi^+C\rightarrow \pi^+ C)$, which follows from isospin symmetry.
The NC cross section is  approximately
half as big as the CC  cross section.
The result is  shown in figure \ref{totalNC} with the solid curve
again corresponding  to $\xi=3$. We plotted also three
experimental points  carried at three different energies and targets made of Carbon,
Aluminum and Freon, respectively.
We use Carbon as our reference nucleus
and scale the results for other nuclei by the $A^{2/3}$ rule, as we discussed
earlier.
Rescaling the Aachen and Gargamelle data, we obtain the points in
Fig.\ref{totalNC}. The three points have large errors and are consistent
with the theoretical curves.  As in the CC case, we should mention though
that the  $\xi=3$ cut, was not imposed in these data. If this was done, the data
would had  been considerably reduced.

\section{Conclusions}

We revisited in this article coherent pion production by neutrinos. There are
several reasons for returning to this old topic. First there are
new data which are becoming available and need an explanation. Second,
we clarify the theoretical framework for  the CC and NC cross section
formulas (\ref{dsigma-full})
and (\ref{dsigma-NC-full}). In doing this,
we decompose the leptonic tensor into density matrix elements
keeping the muon mass. Then we showed that a careful application of PCAC
leads to the formulas (\ref{dsigma-full})  and (\ref{dsigma-NC-full}),
where the bulk of the coherent neutrino pion production is described by the
coherent $\pi N\to \pi N$ scattering, provided $Q^2$ is sufficiently small
and $\nu\gg \sqrt{Q^2}$. Only for such $\nu$-values, we obtain
the simple pion dominating picture presented in this paper.

  A third contribution is the discussion of
data and estimates for the cross sections. To this end
we collected data for the coherent production of pions
on nuclei with pion and photon incident beams. The relevant results for pion
coherent scattering are shown in Fig. \ref{pioncrs}; while  the photoproduction
contributions turned out to be very small. Collecting all terms together we
computed the differential and integrated cross sections shown in
Fig. \ref{diff} -- \ref{totalNC}, keeping the exact phase space.

The results in Fig. \ref{diff} demonstrate that a careful test of PCAC
demands that we keep the muon mass terms and the correct phase space,
because, by  neglecting the muon mass the integrated charged  current cross
section is  overestimated by a factor of two. Corrections from the muon mass
are  also discussed in references \cite{Paschos:2005km, Rein:2006di}
and in the Adler recollections \cite{Adler:1964yx}.
 It will be important for future experiments that a cut like $\xi=3$
is imposed, because only then is the validity of a PCAC treatment
guaranteed. In any case, the integrated cross sections
shown in Figures \ref{totalCC} and \ref{totalNC} are in
satisfactory agreement with experimental data, in view of the large experimental errors.

We feel that the analysis proposed in this article,
together with the use of hadronic data, should provide accurate estimates for
coherent pion production also at higher neutrino energies.

\begin{acknowledgments}
EAP wishes to thank Dr. A. Thomas and colleagues for their
hospitality at Jefferson Laboratory where part of the work was done and
Dr. A. Afanasev for stimulating discussions. The financial support of BMBF,
Bonn under contract 05HT 4 PEA/9   is gratefully  acknowledged. AK wishes to
thank  the Graduiertenkolleg 841 ``Physik der Elementarteilchen
an Beschleunigern und im Universum'' at University of Dortmund for financial
support. In addition,  GJG would like to thank the colleagues
of the Physics Department the  the University of
Dortmund for the hospitality they extended to him.
\end{acknowledgments}

\begin{appendix}

\section{Kinematics}

In this Appendix, we give the kinematic limits for the integration
of the differential cross sections in (\ref{dsigma-full}, \ref{dsigma-NC-full}).

The integration is organized by  first performing the    $t$-integral
in the range
\bq
|t_{\rm min}|< -t < 0.05 {\rm GeV^2}~~, \label{t-limits}
\eq
 where
\bq
t_{\rm min}=\frac{(Q^2+m_\pi^2)^2-\Big [\sqrt{\lambda(W^2,-Q^2,M_N^2)}-
\sqrt{\lambda(W^2,m_\pi^2,M_N^2)}\Big ]^2}{4W^2}\simeq
- \left(\frac{Q^2+m_\pi^2}{2\nu}\right)^2~~, \label{tmin}
\eq
where  $\lambda(a,b,c)=a^2+b^2+c^2-2 ab-2ac-2bc$ is used.

 At fixed  $Q^2$ and  $s \equiv (k_1+P)^2=M_N^2+2M_NE_1$,
the kinematical minimum  and maximum  $\nu$-values are
\begin{align}
\label{numin}
\nu_{min}&=\frac{(W^2_{min}+Q^2-M_N^2)}{ 2M_N} ~~,\\
\label{numax}
\nu_{max}&=\frac{(W^2_{max}+Q^2-M_N^2)}{2M_N}~~,
\end{align}
where
\begin{align}
\label{Wmin}
W^2_{min}&=(M_N+m_\pi)^2~~, \\
W^2_{max}&=\left \{ \frac14s^2\left(1-\frac{M_N^2}{s}\right)^2\left(1-\frac{m_\mu^2}{s}\right)
-\left [ Q^2-\frac{s}{2}\left(1-\frac{M_N^2}{s}\right)
+\frac{m_\mu^2}{2}\left(1+\frac{M_N^2}{s}\right) \right ]^2\right \} \nonumber \\
&\times \left(1-\frac{M_N^2}{s}\right)^{-1}\left(Q^2+m_\mu^2\right)^{-1}~~.
\label{Wmax}
\end{align}
To assure though that the condition for the validity of (\ref{ap-helicity0})
is also satisfied, the $\nu$-integration  is done in the range
\bq
\max \left (\xi \sqrt{Q^2} ~,~ \nu_{min} \right ) <\nu <\nu_{max}~~,
\label{nu-limits}
\eq
where for the present application we selected   $\xi=3$.

Finally, the kinematically allowed minimum $Q^2$ value is
\begin{align}
\label{Qmin}
Q_{min}^2&=\frac{(s-M_N^2)}{2}\left[1-\lambda^\frac12\left(1,
\frac{m_\mu^2}{s},\frac{W^2_{min}}{s}\right)\right]
-\frac12\left[W^2_{min}+m_\mu^2-\frac{M_N^2}{s}(W^2_{min}-m_\mu^2)\right]\\
\end{align}
where (\ref{Wmin}) is used. The interesting $Q^2$-region
for coherent scattering then is
\bq
Q^2_{min} < Q^2 \lsim 0.2~ {\rm GeV^2}~. \label{Q2-limits}
\eq

\end{appendix}

\end{document}